\documentclass{aastex62}

\received{July 5, 2020}
\revised{May 4, 2021}
\accepted{May 5, 2021}

\shorttitle{Properties of GRS 1915+105 $\theta$ Class}
\shortauthors{Banerjee et al.}

\begin{document}

\title{Accretion Flow Properties of GRS 1915+105 During Its $\theta$ Class Using AstroSat Data}

\correspondingauthor{Ayan Bhattacharjee}
\email{ayan@unist.ac.kr}

\author[0000-0001-7796-8907]{Anuvab Banerjee}
\affil{S. N. Bose National Centre for Basic Sciences, Salt Lake, Kolkata 700106, India}

\author[0000-0002-2878-4025]{Ayan Bhattacharjee}
\affiliation{S. N. Bose National Centre for Basic Sciences, Salt Lake, Kolkata 700106, India}
\affiliation{Ulsan National Institute of Science and Technology, Ulsan 44919, Republic of Korea}

\author[0000-0001-6770-8351]{Debjit Chatterjee}
\affiliation{Indian Center for Space Physics, 43 Chalantika, Garia St. Road, Kolkata 700084, India}
\affiliation{Indian Institute of Astrophysics, Koramangala, Bangalore 560034, India}

\author[0000-0003-1856-5504]{Dipak Debnath}
\affiliation{Indian Center for Space Physics, 43 Chalantika, Garia St. Road, Kolkata 700084, India}

\author[0000-0002-0193-1136]{Sandip Kumar Chakrabarti}
\affiliation{Indian Center for Space Physics, 43 Chalantika, Garia St. Road, Kolkata 700084, India}

\author[0000-0002-9418-4001]{Tilak Katoch}
\affiliation{Tata Institute of Fundamental Research, Homi Bhabha Road, Mumbai 400005, India}

\author[0000-0001-7549-9684]{H. M. Antia}
\affiliation{Tata Institute of Fundamental Research, Homi Bhabha Road, Mumbai 400005, India}

\begin{abstract}

The Galactic microquasar GRS 1915+105 shows rich variability which is categorized into different classes. In this paper, 
we report temporal and spectral analysis of GRS 1915+105 to study the properties of the accretion flow when the light curve is showing $\theta$ class variability. For this purpose, we use the Large Area X-ray Proportional Counter (LAXPC) data from the Target of Opportunity observations of India's first multi-wavelength astronomy satellite AstroSat. The $\theta$ class is marked by the recurrent appearance of U-shaped regions in the light curve, where the photon count rate first decreases rapidly and then increases slowly. For our analysis, we use U-shaped regions of the first two orbits (02345 and 02346) on 
2016 March 04. In both of the cases, the dynamic Power Density Spectra (PDS) showed significant power at around $4-5$~Hz, 
suggesting the presence of a low-frequency Quasi-Periodic Oscillation (QPO) around that frequency interval. The QPO frequency is found to increase with time when the energy flux is also enhanced. From the evolution of the spectra, 
we determine the evolution of the accretion flow parameters in both of these observations. Fitting the spectra with the transonic flow solution based Two-Component Advective Flow (TCAF) model in $4-25$~keV energy band shows that the Keplerian disk accretion rate rises with the rise in the radiation intensity, while the location of the centrifugal pressure driven shock front decreases. In both these data, a gradual increment of power-law photon index with intensity is observed suggesting the progressive softening of the source.

\end{abstract}


\section{Introduction} \label{sec:intro}

Since the first discovery by the WATCH detector in 1993 (Castro-Tirado et al.~1992), the Galactic microquasar GRS 1915+105 
aroused a great deal of interest in the scientific community because of its rich and enigmatic spectral and timing variability features. Especially in X-rays, the variation of the luminosity by a few orders of magnitude within a span of a few tens of seconds makes it to be an object of great scientific interest to undertake spectral and temporal study in X-ray band. 
Soon after the discovery, the entire variability of the source was categorized into 14 different variability classes on the 
basis of variability pattern, conventional Color-Color Diagram (CCD) and Hardness Ratio (Belloni et al.~2000; Klein-Wolt 
et al.~2002). The observed recurrence of these classes was a unique feature of this source.\par

Phenomenologically, the X-ray spectra of most X-Ray Binaries (XRBs) can be explained using a thermal and a non-thermal 
component. The thermal component is contributed by the Keplerian accretion disk, emitting at different temperatures from successive concentric shells (Shakura \& Sunyaev 1973). The non-thermal part is generated because of the inverse 
Comptonization of soft disk photons in the coronal region around the black hole (Sunyaev \& Titarchuk 1980; 
McClintock \& Remillard 2006), which may be called a Compton cloud. The spectral and temporal study of transient, variable and persistent sources is necessary to obtain a comprehensive picture of accretion processes around such objects. It was suggested by Belloni et al.~(2000) that in the case of GRS 1915+105, the transition between all the different variability classes \textbf{may be due} to the transition between three basic states --- A, B and C. The states A and B are identified 
by spectral softness, the only difference being the inner disk temperature of B is comparatively higher than that of A. 
In the case of C, the spectrum is dominated by a hard power-law component and the inner disk region is practically non-existent. 
The description of the spectral and temporal behavior of GRS 1915+105 as a transition between these three fundamental 
classes is a unique feature of this object itself and is markedly different from the standard transient source behaviors 
(Tanaka \& Lewin 1995). Qualitative physical pictures of these states and other possible states of the variability class transitions were presented immediately (Chakrabarti \& Nandi, 2000; Nandi, Manickam \& Chakrabarti, 2000).
\par 
The state C or harder state is marked out by the occurrence of 1--15~Hz QPOs. \textbf{Investigating the origin} of 1-15 Hz QPO of GRS 1915+105, Markwardt et al. (1999) reported that QPOs are predominantly present during spectrally hard dips in the \textbf{flaring states and the frequency} of oscillation is strongly correlated with the thermal flux. As the power-law flux is reduced and the spectral slope is steepened towards the end of the dips, the QPO disappears, suggesting the presence of two spectral components in interaction. Chakrabarti \& Manickam (2000) showed that the QPO frequency can be explained by the radial oscillation of the post-shock region, and the transition from soft X-ray domain (burst-on states) to hard dips \textbf{(burst-off states) and the mean} time of residence in those states could be explained by \textbf{outflows} from the post-shock region. They further observed that QPOs are suppressed for thermal photons (0-4 keV), but become prominent for non-thermal photons (4-13 keV), clearly indicating that the \textbf{QPO phenomenon arises from} the Comptonized photons (Rao et al. 2000). Investigation on the energy dependence of low-frequency QPOs of GRS 1915+105 using RXTE data further establishes the inverse Comptonization of the soft seed photons by the `corona' participating in the QPO activity and its frequency variation is strongly correlated with soft X-ray flux (2-5 keV) (Rodriguez et al., 2002). However, it was observed during the analysis of $\alpha$ and $\beta$ classes of GRS 1915+105 that QPO frequency changes during the passage of hard dips, where the soft X-ray flux remains roughly unchanged (Mikles, Eikenberry \& Rothstein, 2006). This further demonstrates the broad correlation of QPO frequency and power-law flux indicated by Markwardt et al. (1999) as well, which gradually weakens towards the latter half of the dip. It had, however, been argued that the two spectral components cannot really be disentangled. This was further supported by spectro-temporal analysis on GRS 1915+105 by Rodriguez et al. (2008) to study the energy dependence of Low-Frequency QPO (LFQPO) in the hard intermediate state. The importance of both of the components had also been highlighted during energy-resolved LFQPO analysis to pinpoint the origin of QPO as well (Vadawale et al.~2001; Rao et al.~2000).
 \par 
Several attempts were made by different groups to determine the origin of QPOs in general. The proposed mechanisms include 
relativistic precessions of encircling matter due to misalignment of the black hole spin axis and inner accretion flow region 
(Stella \& Vietri 1998; Ingram et al.~2009) as well as instability in the inner disk region (Tagger \& Pellat 1999). 
Fitting the phase-resolved spectra using Comptonization, edge smearing and Gaussian equivalent width model, 
Ingram \& van der Klis (2015) suggested the geometric origin of the QPOs.\par  

In the two-component advective flow (TCAF) model, the angular momentum of the inflow is redistributed to form two components: 
 (i) the higher viscosity Keplerian disk on the equatorial plane surrounded by  (ii) the lower viscous and lower angular momentum flow (Chakrabarti 1995; Chakrabarti \& Titarchuk 1995, hereafter CT95; Chakrabarti 1996). It was shown by Chakrabarti and his collaborators that the resonance oscillation of the shocked surface can produce the observed low-frequency QPOs (Molteni et al. 1996; Chakrabarti et al. 2015). Soft seed photons from the Keplerian disk, intercepted by the shock are inverse-Comptonized by  hot electrons in the post-shock region, technically known as the CENtrifugal pressure supported BOundary Layer (CENBOL) (Chakrabarti 1996, 1997; for the nature of different spectra and additional references, see, CT95). More recently, the stability of the TCAF solution was established (Giri \& Chakrabarti 2013) and the Monte-Carlo simulation generated QPOs and observed ones were demonstrated to be similar (Garain et al.~2014). If the cooling time-scale in the CENBOL roughly matches with the infall (i.e., compressional heating) time-scale from the shock front, the resonance oscillation of the shock-surface is triggered. The frequency of oscillation depends on the shock distance from the black hole and the shock strength (Chakrabarti \& Manickam 2000), which in turn, is determined by the viscosity in the equatorial plane. 
The enhancement of viscosity prompts the conversion from the sub-Keplerian to the Keplerian flow, which enhances soft photons cooling the CENBOL. Since the QPO frequency is inversely proportional to the infall time-scale, the frequency increases with the increase of the Keplerian disk rate which controls the cooling (Mondal et al.~2015). During the rising phase of an outburst, 
the shock gradually moves towards the black hole and the QPO frequency steadily increases. Exactly the opposite trend is observed in case of the declining phases, especially in Hard State (HS) and Hard Intermediate State (HIMS) (Debnath et al.~2013). 
To directly fit black hole energy spectra with that from the TCAF model and extract physical flow parameters, this model has been implemented in 2014 as a local additive table model in XSPEC (Debnath et al.~2014, 2015a). After that many transient, 
persistent sources were studied with this model to understand the accretion dynamics of those objects (Debnath et al.~2014, 
2015a,b, 2017, 2020; Mondal et al.~2014, 2016; Jana et al.~2016, 2020a, 2020b; Chatterjee et al.~2016, 2019, 2020a, 2020b; 
Bhattacharjee et al.~2017; Shang et al.~2019, Banerjee et al. 2021). Masses of unknown black hole candidates were also estimated with better accuracies based on the constant model normalization method as described by Molla et al.~(2016). Estimation of X-ray fluxes of the jet and its properties were studied for a few short orbital periods in case of black hole X-ray binaries using a method described by Jana et al.~(2017). The TCAF solution was successfully applied to study the spectral and timing properties of accretion around neutron stars (Bhattacharjee and Chakrabarti 2017; Bhattacharjee 2018; Bhattacharjee and Chakrabarti 2019, 2020, 2021).
\par   
Satellites with broadband instruments play an important role in providing an opportunity to perform both the 
spectral and the temporal analysis over a wide band of wavelengths to derive a complete scientific understanding of such accretion flow mechanisms around accreting objects. The Indian satellite \textit{AstroSat} offers such a platform for 
simultaneous observation in 0.3-100 keV owing to its co-aligned instruments: Large Area X-ray Proportional Counter (LAXPC) 
(Yadav et al.~2016a,b; Antia et al.~2017), Soft X-ray Telescope (SXT) (Singh et al.~2017) and Cadmium Zinc Telluride Imager 
(CZTI) (Vadawale et al.~2015). \textbf{Banerjee et al. (2020) carried out spectral analyses of GRS 1915+105 in its $\chi$ class to determine the physical properties of its radio-loud and radio-quiet sub-classes.} In the present paper, we study another important variable class of the source, namely $\theta$ under TCAF paradigm. \par 

We report the spectral and timing analysis on the $\theta$ class data from a small subset of \textit{AstroSat} Target of Opportunity (ToO) observation on March 4, 2016 by onboard LAXPC instrument. We investigate the evolution of the spectral and temporal behavior and the QPOs (if any), and derive the underlying accretion flow parameters causing the observed evolution. The organization of the paper is as follows: in the next section, we present the details of our observation and data reduction techniques. The detailed results of timing and spectral analysis have been provided in Section 4. 
Finally, in Section 5, we draw the final conclusions.

\section{OBSERVATION AND REDUCTION OF SCIENTIFIC DATA}
The capability of detailed spectro-temporal study of LAXPC has already been demonstrated for several X-ray binaries 
(Yadav et al.~2016a, Misra et al.~2017, Pahari et al.~2017, 2018).  LAXPC instrument consists of three identical but independent X-ray 
Proportional Counters which can register photons in 3--80 keV with a time resolution of $10 \mu$s (Yadav et al.~2016a, 2016b; 
Agrawal et al.~2017). We have carried out our analysis on the basis of instructions provided in the LAXPC analysis software 
$\mathtt{LaxpcSoft}$ released on May 16, 2019\footnote{\url{https://www.tifr.res.in/~astrosat\_laxpc/archived.html}}. The details of the data analysis procedure using background and response matrices are provided by Antia et al.~(2017).
 \par 
Notwithstanding the wealth of data of RXTE on GRS which are available for general usage, there are certain advantages in using AstroSat data. The effective area of RXTE PCA falls quickly with increasing energies. Around 30 keV, the effective area of LAXPC ($\sim$ 4500 $\text{cm}^2$) is significantly higher as compared to RXTE PCA ($\sim$ 1000 $\text{cm}^2$) (Yadav et al. 2016b, Table 1). Further, the time resolution of the standard 2 mode data in RXTE (used for spectral analysis) is 16 seconds, while LAXPC gathers event mode data at high time resolution ($10 \mu$s) which is very useful in the analysis of highly variable sources. For the purpose of our analysis, ToO level 1 data was obtained from the ISSDC data distribution archive 
\footnote{\url{https://astrobrowse.issdc.gov.in/astro\_archive/archive/Home.jsp}}. Only the data from LAXPC10 among the three 
LAXPC detectors were used in our analysis. For the spectral analysis, the source and background spectra have been generated. 
The corrections for South Atlantic Anomaly (SAA), Earth occultation and dead time are done in LAXPC software. In order to avoid the undesirable contribution from the instrumental artifact, namely, a bump at $\sim 33$ keV (Antia et al.~2017), 
only 4.0-25.0 keV energy range is considered. Data from all the anodes in the detector have been used for analysis. 
For the spectral analysis, $\mathtt{XSPEC~v~12.9.1}$ was used.
\par 
In the entire $\sim 67$~ks of LAXPC10 exposure, the data were recorded during a span of nine satellite orbits (02345--02354) 
starting from 
04 March 2016 11:22:15 to 05 March 2016 04:54:23. All the lightcurves in $\theta$ class (Belloni et al.~2000) repeatedly exhibit a `M' shape of a few hundred seconds duration with brief intervals of low count rate. We have selected `U'-shaped 
regions in the `M' shaped $\theta$ class data of the initial two AstroSat orbits, i.e., 02345 and 02346 
(Fig. 1(a,b)). In the orbit no.~02345, this phase spans over 460s (Fig. 1(a)), and in the following orbit, this steady rising phase spans over a wider time period of $\sim 800$ s (Fig. 1(b)). The start times of 
observation in the first and the following orbits correspond to UT 12:01:26 and UT 13:42:14 on the 04 March 2016 
(MJD = 57451.501 and 57451.571 respectively). In our analysis we treat these times as $t_0 = 0$.
\par 
For the first orbit, the photon flux in the entire interval varied from $\sim 2700$~cnt/s to $\sim 7000$~cnt/s 
(Fig 1(a)). In the case of the second orbit, the same thing varies from $\sim 2000$~cnt/s to $\sim 9000$~cnt/s 
(Fig 1(b)). In order to search for any continuous variation of the timing behavior, we adopted a dynamic method for the timing analysis. The individual power-density spectrum (PDS) in the 4--25~keV for one hundred second time 
span is generated using the standard HEASoft XRONOS package task ``powspec'' with a suitable normalization (equals to $-2$) 
to eliminate the white-noise level. For the purpose of analysis, we extracted lightcurves of 0.01 second time resolution, which corresponds to the Nyquist frequency of $50$~Hz. Each lightcurve was divided into $4096$ time bin intervals, and then PDS was created for each interval. Final PDS was obtained by averaging over all these individual PDS. 
The average PDS was binned with bin size increasing by a factor of $1.03$ in frequency space. In the generation of dynamic PDS, each successive PDS is shifted by $1$ second from its previous one. The PDSs are plotted in 3D color map 
with a suitable color palette.
 \par 
In the dynamic PDS, a significant concentration of power is observed within a narrow frequency band for a certain span of observation. Individual PDS were extracted with suitable time resolution and time bin intervals in order to search for the plausible peaked component on top of the broadband noise. The details are provided in Section 3.2. The PDS were fitted using Lorentzian and power-law components in order to take care of the QPO and the broadband component of the noise respectively. Specific attributes corresponding to the peaked component, viz. the centroid frequency, Q factor and contributed rms power were determined from the fitted parameters. The error values were determined 
using the built-in `fit err' method.
  \par 

Within the region of our interest in both orbits, the photon count increased significantly. Therefore, the spectra covering the entire energy domain are not analyzed as a whole. Instead, the entire domain is split into sub-domains depending on the intensity variation. In the case of the first orbit, the domain of analysis is split into five sub-domains: the first one is the 60 seconds interval containing the declining phase, and the other four are 100 seconds intervals in the subsequent rising phase of 400 seconds. In the case of the second orbit, the rising phase of $\sim 700$ seconds is split in equal sub-intervals of 100 seconds each. Spectral analysis has been carried out to obtain the evolution of parameters and in light of this, to understand the timing result. During the entire span of observation, the hydrogen column density 
($N_H$) is kept fixed at $6.0\times 10^{22}~\mathrm{cm}^{-2}$ (Muno et al.~1999) and is modeled using the absorption model 
\textit{phabs}. In order to obtain a quantitative measure of the shift in spectral hardness, each background subtracted 
spectrum is fitted using \textit{phabs*(diskbb+power-law)} model. Subsequently, in order to have a physical picture of the 
accretion flow, we analyze the same spectra using \textit{phabs*TCAF} as well. For this purpose, TCAF based model 
\textit{fits} file is used (\texttt{TCAF\_v0.3.2\_R1.fits}). Four flow parameters (a--d) and one system parameter 
(e) were used to achieve the best fit: (a) Keplerian disk accretion rate, $\dot{m}_d$ (in units of $\dot{M}_\mathrm{Edd}$), 
(b) sub-Keplerian halo accretion rate, $\dot{m}_h$ (in units of $\dot{M}_\mathrm{Edd}$), (c) location of the shock front 
$X_s$ (in unit of Schwarzschild radius $r_g = 2GM/c^2$), (d) the compression ratio of the shock $R$ and (e) mass of the 
black hole (in unit of solar mass $M_\odot$). In order to obtain the best fit, we used a broad Gaussian profile as well to take care of Iron line emission. A $2\%$ systematic error is applied for achieving the best fit (see, Pahari et al.~2017), 
and XSPEC command `err' is used to obtain $90\%$ confidence error values for the spectral model fitted parameters.

\section{RESULTS}
\subsection{Nature of the Lightcurve}
The photon count lightcurves in orbit nos.~02345 and 02346 are shown in Figures 1(a) and 1(b) respectively, where the domains of our timing analysis are clearly marked. In the case of the first orbit, the left-hand side of the vertical line indicates the domain within which we have done timing analysis. For $\sim 60$ seconds after the beginning of the observation, the photon count rate decreases. Subsequently, for $\sim 100$ seconds, the count rate remains almost steady, 
and then it monotonically increases for $\sim 300$ seconds. In the following orbit, we have chosen only the declining and subsequent rising phase from the middle segment of the observation. Within a span of $\sim 700$ seconds, the photon 
count rate varied between $\sim 2000$ and $\sim7000$ cnt/s. 
\par 

\subsection{Timing Analysis}
For the timing analysis, we have generated the dynamic PDS for different domains of observation in both of the orbits. 
The dynamic PDS has been generated by staggering individual PDS of 100 seconds each, such that the origin of each successive time-series advances by one second. In Fig. 2(a), we show the dynamic PDS for the first 460 seconds of 
observation in the orbit no.~02345 to investigate the presence of some frequency domain where the power is concentrated. 
The $x$-labels stand for the mid-point of the respective time windows chosen for the generation of dynamic PDS. 
We observe two modes of oscillation contributing significantly to the overall power. The strongest mode of oscillation is observed at around 4--5~Hz, which is the key focus of our paper, though there is another low-powered and lower frequency mode of oscillation at $\sim 2$~Hz in the first $\sim 250$ seconds as well. With the passage of time, 
we observe a slight drift in this domain of concentrated power. The dynamic PDS generated using the same procedure from the data marked in Fig. 1b, is shown in Fig. 2b. The upward trend in the power concentrated domain is apparent here as well.
\par 

In order to observe the behavior of the peaked componens in the PDS during the sharp declining phases of the two orbits more closely, we generated the dynamic PDS corresponding to that domain separately. The presence of only 60 seconds long duration of the declining phase in case of orbit no. 02345 and sharp decline within $\sim 100$ seconds in case of the 
the following orbit implies that PDS has to be produced for smaller time windows to obtain reasonable PDS that is representative of the declining phase only and is not contaminated by the contribution from the rising phase at all. 
With 100 seconds of data blocks, we pick up noise and peaked components both from the declining and the rising phases. However,  
for smaller data blocks, the corresponding time series needs to be divided into more `newbin intervals' during the production of individual PDS using `powspec' to ensure the pruning of unwanted noise components. Considering these issues, 
the data corresponding to this declining phase was extracted keeping the time resolution of $0.001$ seconds for the purpose of generation of dynamic PDS. Each individual data block was considered to be of 16 seconds duration, and the origin of each successive data-segment was advanced by 1 second. In order to ensure that the final PDS generated on a 
single frame is averaged over four sub-intervals, the time resolution and the number of newbins per sub-interval are kept to be respectively 0.008 second and 512. Fig. 3(a,b) features the dynamic PDS for the declining arm corresponding to the two orbits 02345 and 02346 respectively. For 02345, the power is concentrated between 4-5 Hz for the entire phase. 
In the case of 02346, however, we observe a conspicuous downward trend in the peaked component in frequency space.

We investigated individual PDS to search for the presence of a strong peaked component over the broadband noise. In the case of orbit no. 02345, a closer look on the individual PDS yields the presence of such peaked components in the first 225 seconds of observation, starting from the onset of the rising arm of the U-shaped curve of our interest. In the case of the following orbit, the peaked component is observed in the first 400 seconds of observation during the monotonic increase of the photon flux in the rising arm. Peaked components are also detected during the initial declining phase in both the orbits. 
The PDS are fitted with the combined Lorentzian and power-law models to account for the peaked (QPO) component and broadband noise respectively. The corresponding parameters quantifying the peaks, namely the centroid frequency, the rms power and 
Q-factor (= centroid frequency/FWHM) are determined. In the case of the first orbit, during the declining phase the centroid frequency was found to be around 4.3 Hz, with rms power varying in the range 7.5\% - 9.2\%. Q factor was found to be more than 5 throughout. During the rising phase, centroid frequency drifted from 4.22 Hz to 4.51 Hz. The Q factor was observed to be varying between 7.6 and 12.7. This high Q factor indicates that those peaks can be qualified as QPOs. 
The RMS power dropped from 7.3\% to 5.6\%. Details of the analysis are provided in Table 1. 
A typical Lorentzian model fit around QPO frequency is shown in Fig. 4(Left). 

We fit the lower frequency QPOs using Lorentzian profiles as well. 
The centroid frequency varied within the range   2.05 Hz - 2.13 Hz, 
with the Q-factor in the range 6.5 - 7.0. The r.m.s. power contribution was found to be 
$\sim$ 7\%. Apart from the principal contribution at around 4 Hz, the presence of another mode of oscillation at lower frequency is another feature of our observation. 

The 4 Hz QPO does not seem to be the harmonic of 2 Hz QPO. The 4 Hz QPO is more prominent, sharper and the normalization is higher. The dynamic PDS also suggests that the dominant concentration of power is around 4 Hz. However, detailed 
exploration regarding the characteristics and the physical origin behind the subdominant peaked components merits a full 
length manuscript altogether, and falls outside the scope of the current manuscript.

In the following orbit, the QPO frequency during the declining phase is reduced from $5.39$ Hz to $2.90$ Hz in a span of 
$\sim$ $100$ seconds. Q value resides between 3.8 and 7.9 during this period, with rms power varying from 
6.51\% to 11.87\%. In the rising arm, on the other hand, centroid frequency increases from 2.90 Hz to 4.33 Hz during the domain of investigation. We did not obtain any consistent trend in the Q value or rms power. The Q factor resides in the range of 5.2 to 8.7 and the rms power is observed to vary between 5.8\% and 11.4\%. The details 
of the parameters, i.e., the QPO frequency, Q-value and rms power are noted in Table 1.

\subsection{Spectral Analysis}
The dynamic spectra for both the orbits in their entire observation span are shown in Fig. 5(a,b). In order to produce the dynamic spectrum, we have produced an individual spectrum using a 25~seconds duration data block and then shifting the data block by 1 second. The constant intensity contours repeatedly showed the same pattern even within the observation in one complete orbit. Hence, in order to extract a quantitative estimate of the dynamics of flow parameters with the 
evolution of time, fitting of individual spectra using empirical and physical models have been done only within the first 
U-shaped curve in the first orbit and the prominent U-shaped curve in the second segment of the second orbit. 

For the orbit no.~02345, for the first $\sim 60$ seconds, the count rate gradually goes down and becomes steady for 
$\sim 100$ seconds. In order to determine the flow parameters in all these different domains, the data have been split into five segments. The first segment comprises of $0-60$~seconds data where the count rate is decreasing. Within 
160--460~seconds interval where the photon count rate is almost monotonically increasing, the data have been split into three segments of $100$~seconds each. The power-law photon index changes from 2.602 to 2.546, indicating a transition to relative hardness within this short interval. Subsequently, the photon index goes up to as high as 3.014, indicating gradual softening throughout this period. In order to obtain a quantitative estimate of this transition in terms of physical parameters like accretion rate and shock location, we further performed the spectral analysis using 
\textit{TCAF+Gaussian} model. The variation of flow parameters in \textit{TCAF+Gaussian} model in 4--25~keV, namely 
disk rate ($\dot{m}_d$), halo rate ($\dot{m}_h$) and shock location ($X_s$) corroborate with the result obtained from 
the empirical results. For the purpose of spectral analysis using TCAF, the mass is kept frozen at $14.0M_\odot$. 
Within the first $60$~seconds, the disk accretion rate goes down from $0.79~\dot{M}_\mathrm{Edd}$ to 
$0.77~\dot{M}_\mathrm{Edd}$, and consequently the shock location also recedes from $33.8~r_s$ to $38.9~r_s$. For the next 
$400$~seconds, the disk accretion rate steadily increases from $0.77~\dot{M}_\mathrm{Edd}$ to $0.84~\dot{M}_\mathrm{Edd}$. 
As expected, the shock location decreases from $38.9~r_s$ to $16.9~r_s$ within this time span. This corresponds to the speed of the shock to be about 2200 m/s. This agrees with our finding of the rapid increase of the accretion rate in the disk. This is possible only if the effect is local and the supply at the outer edge may not be affected. The details of the analysis are provided in Table 2. TCAF model fitted unfolded spectrum for 0--60 seconds of observation is given in 
Fig. 4(Right). In Fig. 6, we show the default 68-90-99 percent confidence contours of TCAF fitted parameters 
$\dot{m}_d$ with $\dot{m}_h$ and $\dot{m}_d$ with $X_s$.

In the case of the second orbit, we have done the spectral fitting only in the U-shaped region where the photon count rate first decreases very rapidly and then increases slowly. The \textit{diskbb+power-law} fit in a span of 500 seconds, shows a monotonic increase in the power-law slope, indicating a plausible transition towards the softer state. The disk rate 
increases from $0.60~\dot{M}_\mathrm{Edd}$ to $0.77~\dot{M}_\mathrm{Edd}$, and consequently the shock location goes down 
from $65.8~r_s$ to $55.7~r_s$ within this span. The corresponding shock velocity turns out to be $\sim 900$ m/s. The details of the fitted parameters are shown in Table 2.

\section{SUMMARY AND CONCLUDING REMARKS}
In this paper, we presented our analysis of the spectral and timing behavior of GRS 1915+105 \textbf{using the TCAF paradigm}. For this 
purpose, the $\theta$ class data of the source as obtained by the LAXPC instrument of AstroSat satellite has been used. 
To the best of our knowledge, this is for the first time the $\theta$ class data of AstroSat is being analyzed with the TCAF paradigm. In this paradigm, different spectral states, as well as the QPOs resulting from resonance oscillation of 
the Compton cloud, arise out of the interplay between two types of accretion rates, namely, the Keplerian disk rate 
($\dot{m}_d$) and the sub-Keplerian halo rate ($\dot{m}_h$). In the sub-Keplerian flow, shocks are formed when the 
Rankine-Hugoniot conditions are satisfied (Chakrabarti 1989, 1996; Chakrabarti \& Das 2004). If the Keplerian disk rate increases, it increases the soft seed photons and the post-shock region (CENBOL, which acts as a `Compton cloud') is cooled down rapidly, causing the shock to proceed towards the black hole. This has indeed been obtained in our spectral analysis. 
In the process, both the compressional heating time scale and the cooling time scale of the CENBOL go down and become comparable (Chakrabarti et al.~2015) triggering shock oscillations which manifest as QPOs in the light curve. We first obtained the state of the system by estimating the photon index using \textit{phabs*(diskbb+power-law)} model. Throughout the span of our analysis, the photon index ($\Gamma$) was found to be above $2.4$, indicating that the source was either in soft intermediate state (SIMS) or soft state (SS). In the case of both of the orbits, $\Gamma$ monotonically increases with the increase of photon flux, implying the transition from SS to SIMS. In the first orbit, $\Gamma$ turned out to make an excursion from $2.5$ to $3.0$ during the 400 seconds span of spectral analysis, indicating the transition of the source from SIMS to SS. The gradual enhancement of disk accretion rate and the consequent decline of shock location were also obtained from TCAF model fitting. The shock location moves from $39 r_s$ to $17 r_s$ while disk accretion rate increases 
from $0.77$~to $0.84$~$\dot{M}_\mathrm{Edd}$. A similar trend was followed in the second orbit as well (Table 2). 
In the two orbits (02345 \& 02346) we have analyzed, within a span of 400--500 seconds, the total flux changes by a factor of 3--5. The corresponding shock velocity turns out to be $\sim 2200$ and $900$~m/s respectively. This is significantly 
larger compared to the shock velocity obtained earlier in the case of transient sources, which reside in the ballpark of 
$10-20$~m/s (Chakrabarti et al.~2008; Nandi et al.~2012; Debnath et al.~2013). This suggests the possibility of the local modulation of accretion rate due to feedback from the outflow (Chakrabart \& Manickam, 2000; Chakrabarti \& Nandi, 2000). 
All these results indicate that GRS 1915+105 went through repeated micro-flares when it is in a $\theta$-class. \par 

As the shock location moves closer and becomes weaker, the QPO frequency increases since the frequency is inversely proportional to the infall time scale from the shock. This behavior is common in the rising phase of transient BHCs 
(see, Debnath et al.~2013 and references therein). Therefore, as the source transits from SIMS to SS, the QPO frequency should increase. This has been corroborated by our timing analysis. As shown in Fig. 2(a,b), the QPO frequency shifts towards higher frequency domain with the evolution of time. Within the first $225$~seconds of observation where QPO is relatively prominent, the QPO frequency moves from $4.22$ Hz to $4.51$~Hz in case of the first orbit. In the case of the second orbit, the QPO frequency monotonically increases from 
2.90 Hz to 4.33 Hz within a span of 400 seconds (Table 1).  During this period, shock location 
also decreases from 65.8 $r_s$ to 55.7 $r_s$ (Table 2). These findings further suggest that an interplay of two accretion rates is at the root of manifestation of the spectral and the timing behaviors. We have further observed the lower frequency QPO at $\sim~2$ Hz in the first orbit. The integrated fractional rms turned out to be $\sim 6\%$.
\par
GRS~1915+105 and IGR~17091-3624 exhibit several classes that can be well differentiated by the Comptonizing efficiency 
(Pal \& Chakrabarti 2015). As the variability class moves from the harder states (HS, HIMS) to the softer (SIMS, SS) states, 
cooling increases as the soft photon number increases. This shrinks the CENBOL, resulting inward movement of the shock to raise frequency of the low-frequency QPOs. It has been shown earlier that in the case of $\chi$ class of GRS 1915+105 where X-ray flux is steady, accretion flow could be explained in the light of TCAF paradigm (Banerjee et al., 2020). In the present paper, 
we demonstrated that even inside a single variable class, there are considerable changes in the shock properties and accretion rates. Inferences about the flow from our analysis agree with the TCAF paradigm which explains the spectral and timing properties of GRS 1915+105 to be due to the interplay between the Keplerian and sub-Keplerian flows, the launching of outflows from the CENBOL region, and the effects of radiative transfer within the sonic sphere of the outflow (Nandi, Manickam and Chakrabarti, 2000). The outflow rate depends upon the spectral state (Chakrabarti, 1999) as the outflow is cooled via inverse Comptonization of the intercepted soft seed photons from the disk. This makes part of the outflow from the \textbf{sonic sphere fall back} upon the accreting flow and modulate the accretion rate locally. \textbf{Garain et al. (2012) used time-dependent Monte Carlo simulations to investigate the effects of Compton cooling, confirming the reduction of outflow rate with increased thermal flux.} This accretion flow feedback could explain the rapid variation in accretion rates and consequent shock location in a few tens to a few hundreds of seconds, with a shock velocity ($\sim$ few hundred m/s) significantly higher as compared to usual transient sources during their outbursts ($\sim$ few tens m/s). \textbf{Fig. 5(a,b) suggests a gradual increase of the Keplerian disk accretion with the gradual rise in spectral softness in the U-shaped domain.} With the increase of the disk rate, the shock location moves inwards, causing the QPO frequency to go upwards. However, as the spectral states grow softer with increasing disk rate, the strength of QPOs gradually diminishes. The upward trend of the dominant power with the photon intensity in Fig. 2(a,b) and an explicit downward trend in Fig. 3(a,b) are in agreement with the aforementioned QPO behavior. This interconnected nature of spectral and timing properties manifested in Fig. 2, 3 and 4 are indicative of both effects being of the same origin as suggested in TCAF paradigm. For the first time, we are able to see how the TCAF fitted parameters such as the accretion rates and the shock location change during rapid intensity variations in the count rate of the $\theta$ class of GRS 1915+105.
 \par 

Theoretical investigation on the behavior of magnetic flux tubes under the influence of various competing forces in TCAF configuration had been carried out by Nandi et al. (2001). It was observed that the collapse of the flux tubes under magnetic tension would be the dominant phenomena in hot plasma (temperature $\ge 10^{10}$K), which would expel the matter in transverse direction. This evacuation of disk matter explains the origin of `baby jets'. However, the estimation of the mass contained in the baby jets was found to be one order of magnitude smaller than the earlier estimation (Mirabel et al 1998), which implies sub-Keplerian flow can contribute to such evacuations as well. This picture is aligned with the qualitative scenario of the ejection of Comptonizing region presented by Vadawale et al. (2001, 2003) to explain the association of large radio flares and soft dips.  The importance of disk-jet interaction and the nature of ejecta in determining the X-ray variability patterns was highlighted by Klein-Wolt et al. (2002) as well. In the present paper, we focus our attention on the U-shaped variability of longer duration ($\sim$ few hundred seconds). We find QPOs during the hard intermediate state (bottom part of the U shaped light curve) and their gradual disappearance in the regions of softer states. This implies that the CENBOL is present in the hard intermediate state, though it may gradually form and weaken inside one U-shaped part of the light curve. Our proposition of the collapse of the base of the outflow and the modulation of the accretion rate due to the return flow, which is a natural consequence of the TCAF paradigm, is one possible way to explain this observation. More extensive spectro-temporal analysis to test this hypothesis for various other classes with different intensity profiles would be presented in upcoming papers.

In principle, all these classes could be described by the TCAF paradigm (see, Chakrabarti \& Nandi, 2000; Nandi, Manickam and Chakrabarti, 2000). However one must quantitatively analyze each of these classes to obtain the true reason behind class transitions. It is also important to distinguish between an outburst, where the accretion rates are enhanced near the outer 
edge of the disk and the present system where mini-outbursts are seen with similar enhancements of rates in a very short time scale, leading to the conclusion that the effect is local and could be triggered by the return outflows. 
This work is in progress and would be reported elsewhere.

\acknowledgments
We thank the anonymous referee for his/her comments which has improved the quality of the manuscript. We acknowledge the strong support from Indian Space Research Organization (ISRO) for successful realization and operation of AstroSat mission. The authors also acknowledge the AstroSat team for the distribution. \textbf{LaxpcSoft software is used for analysis}. A.B. and A.B. acknowledge support fellowship of S. N. Bose National Centre for Basic Sciences, Kolkata, India. D.C. and D.D. acknowledge partial support of the DST/GITA sponsored India-Taiwan collaborative project fund (GITA/DST/TWN/P-76/2017). S.K.C. and D.D. acknowledge partial support from ISRO sponsored RESPOND project (ISRO/RES/2/418/17-18) fund. Research of D.D. and S.K.C. is supported in part by the Higher Education Dept. of the Govt. of West Bengal, India.  T.K. and H.M.A. acknowledge the support of the Department of Atomic Energy, Government of India, under project no.~12-R\&D-TFR-5.02-0200.

\facility{Astrosat}



\begin{figure}
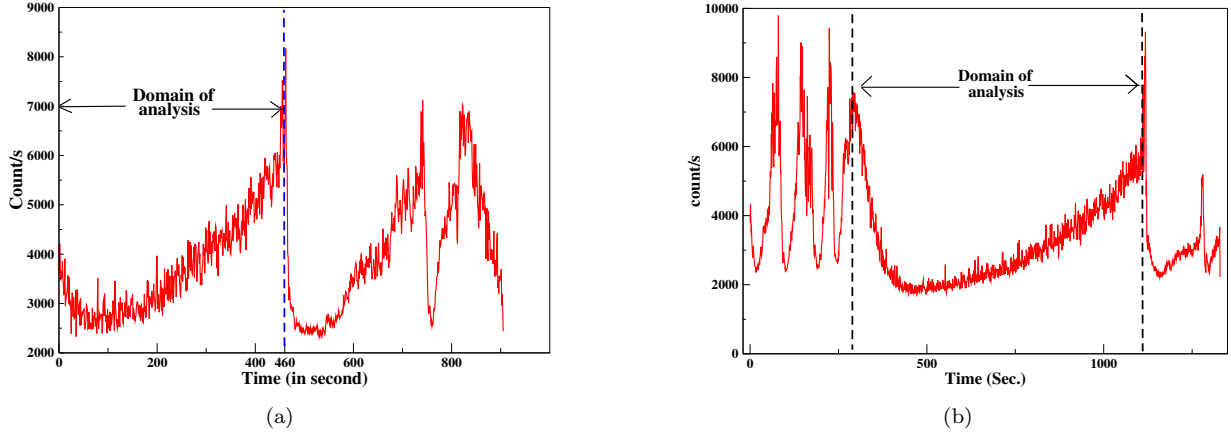

\gridline{\fig{fig1a.eps}{0.4\textwidth}{(a)}
          \fig{fig1b.eps}{0.4\textwidth}{(b)}
          }

\caption{(a) In the left panel, the domain of analysis corresponding to the orbit no.~02345 is indicated. For the purpose of our spectral and timing analysis, we took first 460 sec of data, namely the U-shaped curve of $\theta$ class. in the first $\sim 60$ second the photon count is declining, after which it is monotonically increasing. All the time stamps are relative to $t_0$ (MJD = 57451.501). (b) In the right panel, the domain of analysis corresponding to orbit no.~02346 is indicated. The time stamps are relative to ${t_{0}} = 57451.571.$\label{fig1(a,b)}}
\end{figure}

\begin{figure}
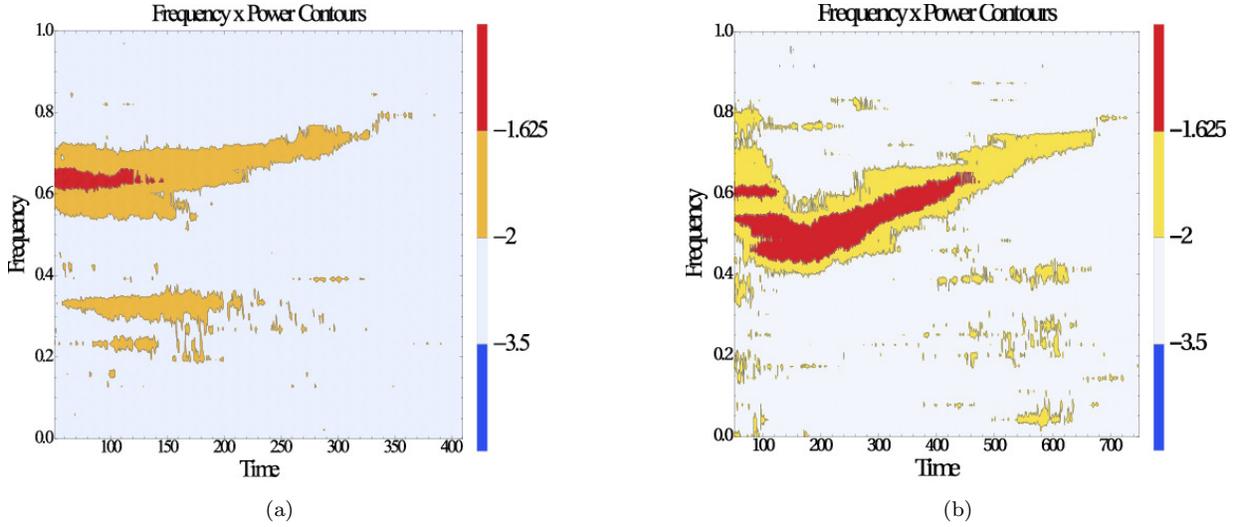

\gridline{\fig{fig2a.eps}{0.4\textwidth}{(a)}
          \fig{fig2b.eps}{0.4\textwidth}{(b)}
          }

\caption{(a) The Power Density Spectra (PDS) for the first 460 Sec. of observation in the orbit no.~02345 is shown. $x$-axis is observation time in second (with $t_0$ at MJD = 57451.501), $y$-axis is Fourier transformed frequency (in Hz) and color bars represent logarithm of frequency multiplied by power. Note, y-axis and color bar is plotted in $\log_{10}$ unit. The strongest mode of oscillation have been found to be around 4 Hz and that mode is conspicuous for about the first 225 Sec. This domain has been chosen for spectral analysis. Apart from that, the low-frequency oscillation at $\sim 2 Hz$ is also obtained. (b) Corresponding to orbit no.~02346, we have shown the dynamic PDS for the entire domain of our analysis as indicated in Figure 1(b). The $0^{th}$ of the time stamps stand for 300 Sec. after the initiation of observation for the orbit no.~02346 (= MJD 57451.571), i.e. MJD 57451.574. The low frequency oscillation mode is absent and the strongest mode of oscillation lies around $3-4$Hz.\label{fig2(a,b)}}
\end{figure}

\begin{figure}
\gridline{\fig{fig3a.eps}{0.4\textwidth}{(a)}
          \fig{fig3b.eps}{0.4\textwidth}{(b)}
          }

\caption{(a) The Power Density Spectra (PDS) for the first 64 seconds of observation corresponding to the orbit no. 02345 is shown. $x$-axis represents the observation time in second (with $t_0$ at MJD = 57451.501), $y$-axis is Fourier transformed frequency (in Hz). Color bar represents frequency multiplied by power. Both y-axis and color bar are provided in $\log_{10}$ unit. The x-labels stand for the mid point of corresponding data-blocks. The strongest mode of oscillation have been found to be within 4-5 Hz. (b) The dynamic PDS corresponding to The fast declining phase in the domain of our analysis as indicated in figure 1(b) is shown. The time stamps are relative to the $t_0$ of orbit 02346 (= MJD 57451.571), i.e. MJD 57451.574. The declining trend in the power concentrated domain from 30 seconds onward is apparent. \label{fig3(a,b)}}
\end{figure}

\begin{figure}
\includegraphics[scale=0.35,angle=270]{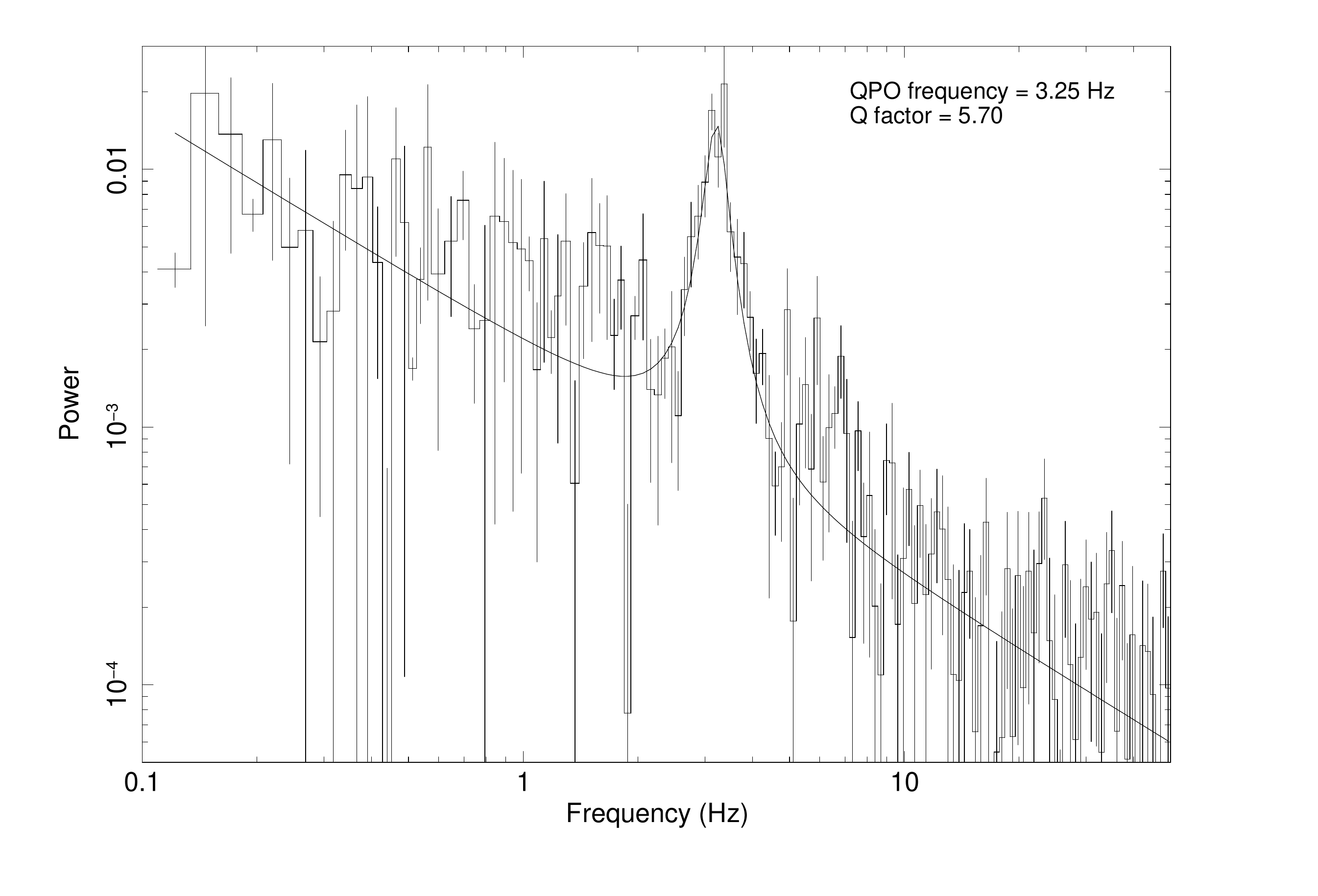}
\hfill
\includegraphics[scale=0.35,angle=270]{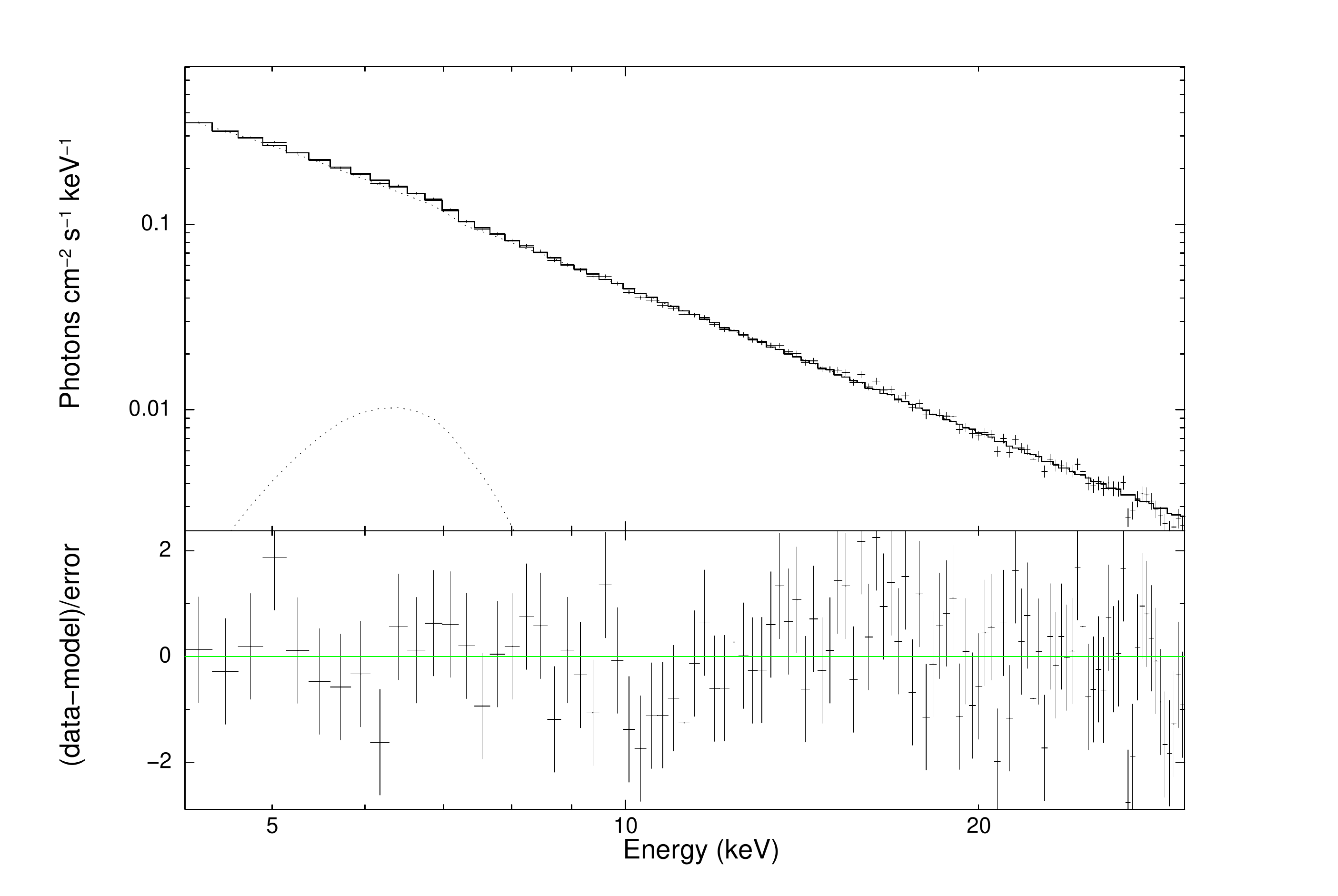}

\caption{(Left) We show the power density spectrum obtained for 500--600 Sec observation in orbit no.~02346. From the Lorentzian fitting, the QPO frequency turned out to be 3.25 Hz and Q factor was found to be 5.70. The RMS power contributed was 10.79$\%$. (Right) The 4--25 keV phabs*(TCAF+Gaussian) fitted unfolded spectrum with residue for the 0--60 second observation for the orbit no. 02345 have been shown.\label{fig3(a,b)}}
\end{figure}

\begin{figure}
\gridline{\fig{fig5a.eps}{0.45\textwidth}{(a)}
          \fig{fig5b.eps}{0.45\textwidth}{(b)}
          }

\caption{(a) The dynamic spectrum for the entire range of the orbit no.~02345 is shown. The contours of constant intensity show similar patterns in different U-shaped branches of $\theta$ class. (b) Dynamic spectra for $\sim$ 1000 seconds starting from the onset of the U-shaped region as indicated in Figure 1(b). Within the domain of analysis, the constant intensity contours indicate the gradual softening of the spectra.\label{fig4(a,b)}}
\end{figure}

\begin{figure}
\includegraphics[scale=0.35,angle=270]{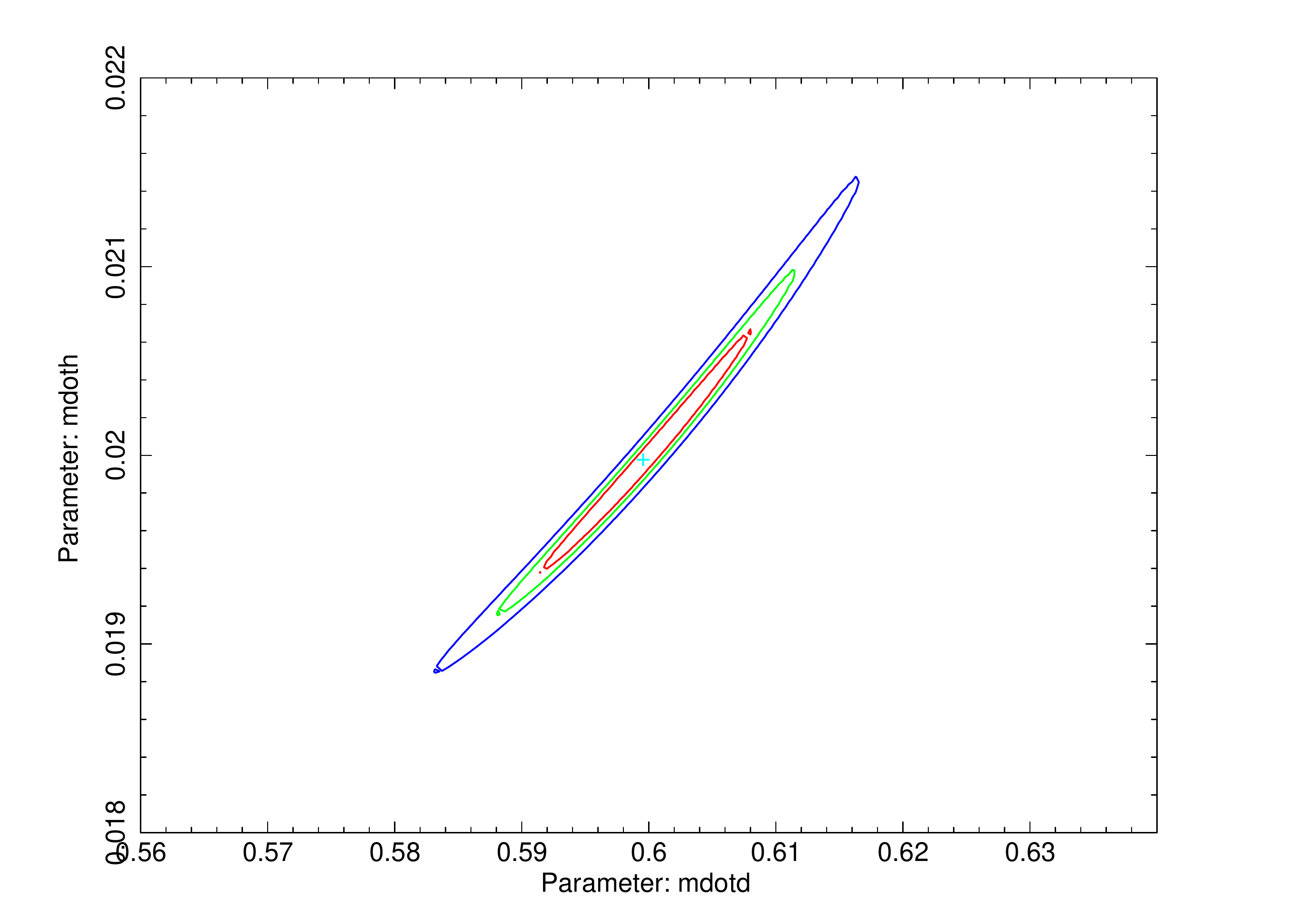}
\hfill
\includegraphics[scale=0.35,angle=270]{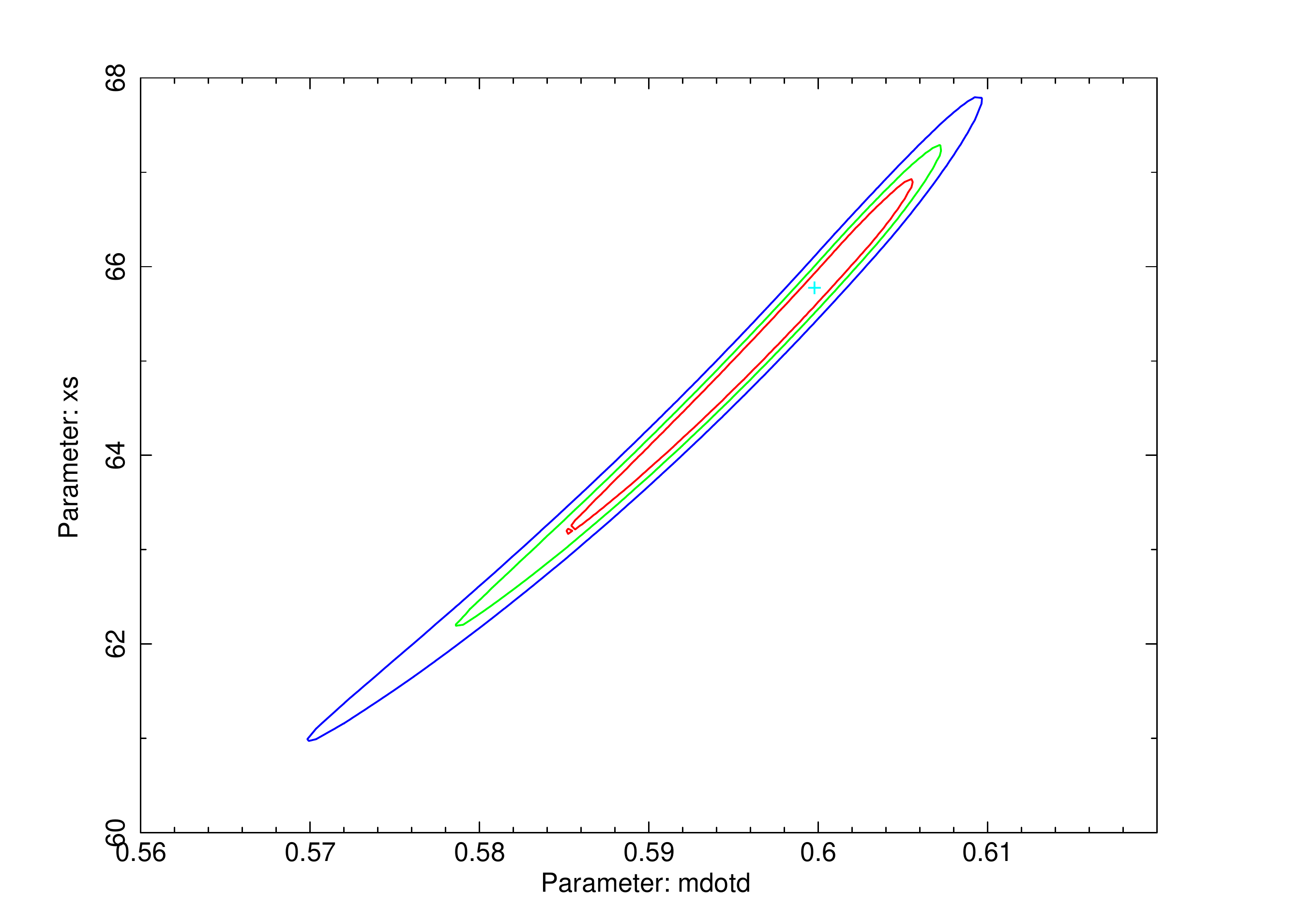}

\caption{(Left) Default 68-90-99 percent confidence contours for TCAF fitted parameters $\dot{m}_d$ and $\dot{m}_h$ are shown for the 400-500 second of observation in orbit no.~02346. (Right) Confidence contours for the same observation shown for $\dot{m}_d$ and $X_s$.\label{fig5(a,b)}}
\end{figure}

\clearpage

\begin{table}[ht]
{\centerline{\large \textbf{Table 1}}}
\vspace{0.2 cm}
\begin{center}
 \begin{tabular}{ c  c  c  c  c} 
 \hline
\textbf{Orbit} & \textbf{Time} & \textbf{QPO} & \textbf{Q} & \textbf{RMS} \\ 
 \textbf{no.} & \textbf{segment} & \textbf{frequency} & \textbf{Value}  \\
& \textbf{(Second)} & \textbf{(Hz)}  &  & (\%)\\
 \hline
  \vspace{0.20cm}                         

& 50-150 \hspace{0.15cm} & $4.22 \pm 0.05$ \hspace{0.15cm} & $10.0 \pm 0.2$ \hspace{0.15cm} &  $7.3 \pm 0.2$ \hspace{0.15cm} 
 \\
 \vspace{0.20cm}                         

02345 & 75-175 \hspace{0.15cm} & $4.27 \pm 0.05$ \hspace{0.15cm} & $12.7 \pm 0.2$ \hspace{0.15cm} &  $6.7 \pm 0.3$ \hspace{0.15cm}
 \\
 \vspace{0.20cm}                         

& 100-200 \hspace{0.15cm} & $4.41 \pm 0.04$ \hspace{0.15cm} & $7.6 \pm 0.1$ \hspace{0.15cm} &  $7.3 \pm 0.2$ \hspace{0.15cm}
 \\
 \vspace{0.20cm}                         

& 125-225 \hspace{0.15cm} & $4.51 \pm 0.09$ \hspace{0.15cm} & $11.3 \pm 0.1$ \hspace{0.15cm} &  $5.6 \pm 0.3$ \hspace{0.15cm}
 \\
\hline

  & 332-348 \hspace{0.15cm} & $5.39 \pm 0.04$ \hspace{0.15cm} & $6.0 \pm 0.1$ \hspace{0.15cm} &  $6.5 \pm 0.3$ \hspace{0.15cm} \\
 \vspace{0.20cm}
 
   & 348-364 \hspace{0.15cm} & $4.87 \pm 0.03$ \hspace{0.15cm} & $3.8 \pm 0.1$ \hspace{0.15cm} &  $8.9 \pm 0.3$ \hspace{0.15cm} \\
 \vspace{0.20cm}
 
    & 364-380 \hspace{0.15cm} & $4.20 \pm 0.04$ \hspace{0.15cm} & $7.9 \pm 0.1$ \hspace{0.15cm} &  $11.5 \pm 0.3$ \hspace{0.15cm} \\
 \vspace{0.20cm}
 
    & 380-396 \hspace{0.15cm} & $3.69 \pm 0.05$ \hspace{0.15cm} & $3.7 \pm 0.2$ \hspace{0.15cm} &  $11.9 \pm 0.2$ \hspace{0.15cm} \\
 \vspace{0.20cm}
 
  & 400-500 \hspace{0.15cm} & $2.90 \pm 0.05$ \hspace{0.15cm} & $7.8 \pm 0.2$ \hspace{0.15cm} &  $9.4 \pm 0.3$ \hspace{0.15cm} \\
 \vspace{0.20cm}
 
 & 425-525 \hspace{0.15cm} & $2.98 \pm 0.03$ \hspace{0.15cm} & $8.3 \pm 0.1$ \hspace{0.15cm} &  $11.4 \pm 0.3$ \hspace{0.15cm} \\
 \vspace{0.20cm}
 
 & 450-550 \hspace{0.15cm} & $3.02 \pm 0.05$ \hspace{0.15cm} & $6.7 \pm 0.2$ \hspace{0.15cm} &  $11.0 \pm 0.3$ \hspace{0.15cm} \\
 \vspace{0.20cm}
 
 & 475-575 \hspace{0.15cm} & $3.11 \pm 0.03$ \hspace{0.15cm} & $8.7 \pm 0.1$ \hspace{0.15cm} &  $9.5 \pm 0.3$ \hspace{0.15cm} \\
 \vspace{0.20cm}
 
&  500-600 \hspace{0.15cm} & $3.25 \pm 0.04$ \hspace{0.15cm} & $5.7 \pm 0.2$ \hspace{0.15cm} &  $10.8 \pm 0.2$ \hspace{0.15cm} \\
 \vspace{0.20cm}
 
02346 & 525-625 \hspace{0.15cm} & $3.38 \pm 0.04$ \hspace{0.15cm} & $6.0 \pm 0.1$ \hspace{0.15cm} &  $9.4 \pm 0.2$ \hspace{0.15cm} \\
 \vspace{0.20cm}
 
 & 550-650 \hspace{0.15cm} & $3.56 \pm 0.05$ \hspace{0.15cm} & $5.9 \pm 0.1$ \hspace{0.15cm} &  $9.2 \pm 0.2$ \hspace{0.15cm} \\
 \vspace{0.20cm}
 
& 575-675 \hspace{0.15cm} & $3.77 \pm 0.06$ \hspace{0.15cm} & $5.7 \pm 0.1$ \hspace{0.15cm} &  $9.2 \pm 0.2$ \hspace{0.15cm} \\
 \vspace{0.20cm}
 
 & 600-700 \hspace{0.15cm} & $3.81 \pm 0.06$ \hspace{0.15cm} & $6.2 \pm 0.1$ \hspace{0.15cm} &  $10.1 \pm 0.3$ \hspace{0.15cm} \\
 \vspace{0.20cm}
 
 & 625-725 \hspace{0.15cm} & $3.92 \pm 0.05$ \hspace{0.15cm} & $6.2 \pm 0.1$ \hspace{0.15cm} &  $9.4 \pm 0.2$ \hspace{0.15cm} \\
 \vspace{0.20cm}
 
 & 650-750 \hspace{0.15cm} & $4.05 \pm 0.07$ \hspace{0.15cm} & $8.6 \pm 0.2$ \hspace{0.15cm} &  $7.7 \pm 0.2$ \hspace{0.15cm} \\
 \vspace{0.20cm}
 
 & 675-775 \hspace{0.15cm} & $4.28 \pm 0.09$ \hspace{0.15cm} & $8.1 \pm 0.1$ \hspace{0.15cm} &  $5.8 \pm 0.3$ \hspace{0.15cm} \\
 \vspace{0.20cm}
 
  & 700-800 \hspace{0.15cm} & $4.33 \pm 0.06$ \hspace{0.15cm} & $5.2 \pm 0.1$ \hspace{0.15cm} &  $6.3 \pm 0.3$ \hspace{0.15cm} \\

  \hline
  
 \vspace{0.20cm}
\end{tabular}
\end{center}
\caption{The variation of QPO frequencies, respective Q-values and RMS powers are shown w.r.t. time. All the time stamps are relative to $t_0$ of the corresponding orbit. Here, 2016 March 4 data of AstroSat orbit number 02345 and 02346 are used. $t_0$ for 02345 is defined at UT 12:01:26, i.e., at MJD=57451.501. For the following orbit, $t_0$ is defined at UT 13:42:14, i.e., at MJD=57451.571. The QPO frequency slowly drifts as time evolves.}

\end{table}


\begin{table}[ht]
{\centerline{\large \textbf{Table 2}}}
\vspace{0.2 cm}
\begin{center}
 \begin{tabular}{c  c  c  c  c  c  c  c } 
 \hline
\textbf{Orbit} &  \textbf{Time} & \textbf{$\dot{m}_d$} & \textbf{$\dot{m}_h$} & \textbf{$X_s$} & \textbf{R} & \textbf{$\Gamma$}  & \textbf{$\chi^2/DOF$}\\ 
\textbf{no.} & \textbf{segment} &  &   &  &  &  &\\
 &\textbf{(Second)} &  ($\dot{M}_{Edd}$)  &  ($\dot{M}_{Edd}$) & $(r_s)$ &  &  & \\
 \hline
  \vspace{0.20cm} 
&  0-60 \hspace{0.15cm} & $0.787_{-0.044}^{+0.047}$ \hspace{0.15cm} & $0.082_{-0.004}^{+0.005}$ \hspace{0.15cm} &  $33.754_{-0.048}^{+0.096}$  \hspace{0.15cm}  &  $1.457_{-0.001}^{+0.002}$ &  $2.602_{-0.044}^{+0.044}$  &  74.59/79\\
\vspace{0.15cm}                        
&  60-160 \hspace{0.15cm} & $0.772_{-0.028}^{+0.027}$ \hspace{0.15cm} & $0.074_{-0.003}^{+0.003}$ \hspace{0.15cm} &  $38.884_{-0.117}^{+0.314}$  \hspace{0.15cm}  &  $1.496_{-0.001}^{+0.002}$ &  $2.546_{-0.038}^{+0.038}$  &  79.09/79\\
\vspace{0.15cm}
02345 &  160-260 \hspace{0.15cm} & $0.779_{-0.053}^{+0.054}$ \hspace{0.15cm} & $0.077_{-0.004}^{+0.001}$ \hspace{0.15cm} &  $33.903_{-0.058}^{+0.140}$  \hspace{0.15cm}  &  $1.487_{-0.001}^{+0.001}$ &  $2.605_{-0.037}^{+0.037}$  &  57.49/79\\
\vspace{0.15cm}

&  260-360 \hspace{0.15cm} & $0.793_{-0.038}^{+0.039}$ \hspace{0.15cm} & $0.108_{-0.002}^{+0.002}$ \hspace{0.15cm} &  $24.151_{-0.064}^{+0.065}$  \hspace{0.15cm}  &  $1.372_{-0.002}^{+0.001}$ &  $2.539_{-0.034}^{+0.034}$  &  58.32/79\\
\vspace{0.15cm}

&  360-460 \hspace{0.15cm} & $0.839_{-0.058}^{+0.058}$ \hspace{0.15cm} & $0.107_{-0.001}^{+0.002}$ \hspace{0.15cm} &  $16.939_{-0.047}^{+0.048}$  \hspace{0.15cm}  &  $1.503_{-0.001}^{+0.001}$ &  $3.014_{-0.033}^{+0.033}$  &  68.11/79\\
 
\hline

&  400-500 \hspace{0.15cm} & $0.599_{-0.033}^{+0.034}$ \hspace{0.15cm} & $0.020_{-0.001}^{+0.002}$ \hspace{0.15cm} &  $65.837_{-0.062}^{+0.061}$  \hspace{0.15cm}  &  $1.222_{-0.002}^{+0.002}$ &  $2.475_{-0.038}^{+0.034}$  &  95.71/79\\
\vspace{0.15cm}
&  500-600 \hspace{0.15cm} & $0.658_{-0.047}^{+0.043}$ \hspace{0.15cm} & $0.020_{-0.002}^{+0.001}$ \hspace{0.15cm} &  $61.390_{-0.072}^{+0.070}$  \hspace{0.15cm}  &  $1.231_{-0.001}^{+0.002}$ &  $2.542_{-0.051}^{+0.055}$  &  103.52/79\\
\vspace{0.15cm}
&  600-700 \hspace{0.15cm} & $0.676_{-0.054}^{+0.055}$ \hspace{0.15cm} & $0.020_{-0.001}^{+0.001}$ \hspace{0.15cm} &  $60.351_{-0.084}^{+0.088}$  \hspace{0.15cm}  &  $1.232_{-0.001}^{+0.001}$ &  $2.516_{-0.047}^{+0.044}$  & 103.97/79\\
\vspace{0.15cm}
02346 &  700-800 \hspace{0.15cm} & $0.750_{-0.062}^{+0.067}$ \hspace{0.15cm} & $0.021_{-0.001}^{+0.001}$ \hspace{0.15cm} &  $58.378_{-0.071}^{+0.076}$  \hspace{0.15cm}  &  $1.264_{-0.001}^{+0.001}$ &  $2.652_{-0.058}^{+0.057}$  &  100.38/79\\
\vspace{0.15cm}
&  800-900 \hspace{0.15cm} & $0.766_{-0.075}^{+0.077}$ \hspace{0.15cm} & $0.022_{-0.001}^{+0.002}$ \hspace{0.15cm} &  $56.031_{-0.081}^{+0.084}$  \hspace{0.15cm}  &  $1.285_{-0.001}^{+0.001}$ &  $2.707_{-0.041}^{+0.039}$  &  75.30/79\\
\vspace{0.15cm}
&  900-1000 \hspace{0.15cm} & $0.767_{-0.045}^{+0.047}$ \hspace{0.15cm} & $0.023_{-0.001}^{+0.001}$ \hspace{0.15cm} &  $55.671_{-0.061}^{+0.067}$  \hspace{0.15cm}  &  $1.543_{-0.001}^{+0.001}$ &  $2.759_{-0.045}^{+0.049}$  &  82.72/79\\
\hline
\end{tabular}
\end{center}
\caption{For the domain of analysis in orbit no.~02345 and 02346, the variation of TCAF fitted parameters, namely disk accretion rate ($\dot{M}_{d}$), halo rate ($\dot{M}_{h}$), shock location ($X_s$), shock strength (R) and power-law photon index obtained from \textit{diskbb + power-law} model w.r.t. time are shown. The mass of the black hole is kept frozen.}

\end{table}




\begin{thebibliography}{99}

\bibitem[Agarwal et al.(2017)]{2017JApA..38..30} Agrawal, P. C., Yadav, J. S., Antia, H. M., et al. 2017, JApA, 38, 30
\bibitem[Antia et al.(2017)]{2017ApJS...231..10} Antia, H. M., Yadav, J. S., Agrawal, P. C., et al. 2017, ApJS, 231, 10
\bibitem[Banerjee et al.(2020)]{Banerjee20}Banerjee, A., Bhattacharjee, A., Debnath, D., Chakrabarti, S. K., 2020, RAA, 20(12), 208 (arXiv:1905.01538v2)
\bibitem[Banerjee et al.(2021)]{B21} Banerjee, I., Bhattacharjee, A., Banerjee, A., et al., 2021, Submitted to RAA, arXiv:1904.11644
\bibitem[Belloni etal.(2000)]{2000A..355..271} Belloni, T., Klein-Wolt, M., Meńdez, M., van der Klis, M., \& van Paradijs, J. 2000, A\&A, 355, 271
\bibitem[Bhattacharjee et al. (2017)]{Bhattacharjee17} Bhattacharjee, A., Banerjee, I., Banerjee, A., et al. 2017, \mnras, 466, 1372
\bibitem[Bhattacharjee \& Chakrabarti (2017)]{BC17} Bhattacharjee, A., \& Chakrabarti, S. K., 2017, \mnras, 472, 1361
\bibitem[Bhattacharjee(2018)]{2018ASSP...53...93B} Bhattacharjee, A.\ 2018, Exploring the Universe: From Near Space to Extra-Galactic, 53, 93. doi:10.1007/978-3-319-94607-8\_8
\bibitem[Bhattacharjee \& Chakrabarti (2019)]{BC19} Bhattacharjee, A., \& Chakrabarti, S. K., 2019, \apj, 873, 119
\bibitem[Bhattacharjee \& Chakrabarti (2020)]{BC20} Bhattacharjee, A., \& Chakrabarti, S. K., 2020, Submitted to JKAS (arXiv: 2012.14502)
\bibitem[Bhattacharjee \& Chakrabarti (2019)]{BC21} Bhattacharjee, A., \& Chakrabarti, S. K., 2021, Submitted to JKAS (arXiv: 2101.01488)
\bibitem[Tirado et al.(1992)]{1992IAU} Castro-Tirado A.J., Brandt S., Lund S., 1992, IAU Circ. 5590
\bibitem[Chakrabarti S. K. (1989)]{ApJ...347...365} Chakrabarti. S. K., 1989, ApJ, 347,365
\bibitem[Chakrabarti (2000)]{ApJ..531..41-44} Chakrabarti, S.K., Manickam, S.G., ApJ 531, 41–44, 2000
\bibitem[Chakrabarti (2004)]{ApJ..349..649} Chakrabarti, S. K. \& Das, S., 2004, MNRAS, 349, 649
\bibitem[Chakrabarti (1995)]{NEWYORK..759..546} Chakrabarti S. K., 1995, in Bohringer H., Morfil G. E., Trumper J., eds, Seventeenth Texas Symposium on Relativistic Astro-physics and Cosmology, Vol. 759. New York Academy of Sciences, New York, p. 546
\bibitem[CT(95)]{ApJ...455..623} Chakrabarti, S., \& Titarchuk, L. G. 1995, ApJ, 455, 623
\bibitem[Chakrabarti (1996)]{ApJ...464..664} Chakrabarti, S. K. 1996, ApJ, 464, 664
\bibitem[Chakrabarti (1997)]{ApJ..484..313} Chakrabarti, S. K. 1997, ApJ, 484, 313
\bibitem[Chakrabarti (1999)]{ApJ..}  Chakrabarti, S. K., 1999,  Astron Astrophys, 351, 185 
\bibitem[Chakrabarti et al.(2008)]{AA..489..L41} Chakrabarti, S. K., Debnath, D., Nandi, A., Pal, P. S., 2008, A\&A, 489, L41-L44 
\bibitem[CM (2000)]{ApJ...531...L41} Chakrabarti S. K., Manickam S. G., 2000, ApJ, 531, L41
\bibitem[Chakrabarti et al.(2015)]{MNRAS..452..3451} Chakrabarti S. K., MOndal, S. \& Debnath, D., 2015, MNRAS, 452, 3451–3456
\bibitem[Nandi et al.(2000)]{astro..ph..0012523} Nandi A., Manickam S. G., Chakrabarti S. K., 2000, preprint (astro-ph/ 0012523)
\bibitem[Chakrabarti (2000)]{IJP..75B..1} Chakrabarti, S.K. \& Nandi,A., 2000, Ind. J. Phys., 75(B), 1 (astro-ph/0012526)
\bibitem[Chatterjee et al. (2016)]{CDC16} Chatterjee, D., Debnath, D., Chakrabarti, S. K., et al., 2016, \apj, 827, 88
\bibitem[Chatterjee et al. (2019)]{CDC19} Chatterjee, D., Debnath, D., Chakrabarti, S. K., Jana, A., 2019, \apss, 364, 14
\bibitem[Chatterjee et al. (2020a)]{CDC20} Chatterjee, K., Debnath, D., Chatterjee, D., Jana, A., Chakrabarti, S. K., 2020a, \mnras, 493, 2452 
\bibitem[Chatterjee et al. (2020b)]{Chatterjee20b} Chatterjee, K., Debnath, D., Banerjee, A., et al., 2020b, RAA (submitted) (arXiv:2006.09077)
\bibitem[Debnath et al.(2013)]{ASR..52..2143} Debnath D., Chakrabarti S. K., Nandi A., 2013, Advances in Space Research, 52, 2143
\bibitem[Debnath et al. (2014)]{DCM14}Debnath D., Chakrabarti S. K., Mondal S., 2014, \mnras, 440, L121
\bibitem[Debnath et al. (2015a)]{DMC15}Debnath D., Mondal S., Chakrabarti S. K., 2015a, \mnras, 447, 1984
\bibitem[Debnath et al. (2015b)]{DMCM15}Debnath D., Molla A. A., Chakrabarti S. K., Mondal S., 2015b, \apj, 803, 59
\bibitem[Debnath et al. (2017)]{DJCCM17}Debnath D., Jana A., Chakrabarti S. K., Chatterjee D., Mondal S., 2017, \apj, 850, 92
\bibitem[Debnath et al. (2020)]{DD20}Debnath D., Chatterjee D., Jana A., Chakrabarti S. K., Chatterjee K., 2020, RAA, 20, 175
\bibitem[Garain et al.(2012)]{ApJ..758..114} Garain S. K., Ghosh H., Chakrabarti S. K., 2014, ApJ, 758, 114
\bibitem[Garain et al.(2014)]{MNRAS...437..1329} Garain S. K., Ghosh H., Chakrabarti S. K., 2014, MNRAS, 437, 1329
\bibitem[Giri (2013)]{MNRAS...430..2836} Giri, K., \& Chakrabarti, S. K. 2013, MNRAS, 430, 2836
\bibitem[Homan et al.(2005)]{ApJ..623..383} Homan J., Miller J. M., Wijnands R., van der Klis M., Belloni T., Steeghs D., Lewin W. H. G., 2005, ApJ, 623, 383
\bibitem[Ingram (2015)]{MNRAS..446..3516} Ingram A., van der Klis M., 2015, MNRAS, 446, 3516
\bibitem[Ingram et al. (2009)]{MNRAS..397..L101} Ingram A., Done C., Fragile P. C., 2009, MNRAS, 397, L101
\bibitem[Jana et al. (2016)]{JDC16}Jana A., Debnath D., Chakrabarti S. K. et al., 2016, \apj, 819, 107
\bibitem[Jana et al. (2017)]{JCD17} Jana, A., Chakrabarti, S. K., \& Debnath, D., 2017, \apj, 850, 91
\bibitem[Jana et al. (2020a)]{Jana20a} Jana, A., Debnath, D., Chakrabarti, S. K., \& Chatterjee, D., 2020a, RAA, 20, 28
\bibitem[Jana et al. (2020a)]{Jana20b} Jana, A., Debnath, D., Chatterjee, D., et al. 2020b, \apj, 897, 3
\bibitem[Klein Wolt et al.(2002)]{MNRAS..331..745} Klein-Wolt, M., Fender, R. P., Pooley, G. G., et al. 2002, MNRAS, 331, 745
\bibitem[Markwardt et al.(1999)]{ApJ...513...L37} Markwardt C. B., Swank J. H., Taam R. E., 1999, ApJ, 513, L37
\bibitem[McClintock (2006)]{Cambridge, ISBN 0521826594} McClintock J. E., Remillard R. A., 2006, in Lewin W., van der Klis M., eds, Black Hole Binaries, Compact Stellar X-ray Sources. Cambridge Univ. Press, Cambridge, ISBN 0521826594
\bibitem[Mikies et al. (2006)]{ApJ..637..978} Mikles, V. J., Eikenberry, S. S., \& Rothstein, D. M. 2006, ApJ, 637, 978
\bibitem[Mirabel et al. (1998)]{AA..330..L9} Mirabel, I. F., Dhaman, V., Chaty, S., et al. 1998, A\&A, 330, L9
\bibitem[Misra et al. (2017)]{ApJ...835..195} Misra, R., Yadav, J. S., Verdhan Chauhan, J., et al. 2017, ApJ, 835, 195
\bibitem[Molla et al. (2016)]{MDC16}Molla A. A., Debnath D., Chakrabarti S. K. et al., 2016, \mnras, 460, 3163
\bibitem[Molla et al. (2017)]{MCDM17}Molla A. A., Chakrabarti S. K., Debnath D., Mondal S., 2017, \apj, 834, 88
\bibitem[Molteni et al. (1996)]{MSC96} Molteni, D., Sponholz, H. \& Chakrabarti, S. K., 1996, ApJ, 457, 805
\bibitem[Mondal et al. (2014)]{MDC14}Mondal S., Debnath D., Chakrabarti S. K., 2014, \apj, 784, 4
\bibitem[Mondal et al.(2015)]{ApJ..798..57} Mondal S., Chakrabarti S. K., Debnath D., 2015, \apj, 798, 57
\bibitem[Mondal et al. (2016)]{MCD16}Mondal S., Chakrabarti S. K., Debnath D., 2016, \apss, 361, 309
\bibitem[Muno et al.(1999)]{ApJ...527..321} Muno, M. P., Morgan, E. H., \& Remillard, R. A. 1999, ApJ, 527, 321
\bibitem[Nandi et al. (2001)]{AA} Nandi, A., Chakrabarti, S. K., Vadawale, S. V. \& Rao, A. R., 2001, A\&A, 380, 245
\bibitem[Nandi et al.(2012)]{AA..542..A56} Nandi, A., Debnath, D., Mandal, S., Chakrabarti, S. K., 2013, A\&A 542, A56
\bibitem[Pahari et al.(2017)]{ApJ...846..16} Pahari, M., Antia, H. M., Yadav, J. S., et al. 2017, ApJ, 849, 16
\bibitem[Pahari et al.(2018)]{ApJL..853..L11} Pahari, M., Yadav, J. S., Verdhan Chauhan, J., et al. 2018, ApJL, 853, L11
\bibitem[Pal et al.(2015)]{ADSPR..56..1784} Pal, P. S., Chakrabarti, S. K., 2015, Advances in Space Research, 56, 1784
\bibitem[Rao et al.(2000)]{AA...360..L25} Rao, A. R., Naik, S., Vadawale, S. V., \& Chakrabarti, S. K., 2000, A\&A, 360, L25
\bibitem[Rodriguez et al. (2002)]{AA...386...271} Rodriguez, J., Durouchoux, Ph.,Mirabel, I. F., Ueda,Y., Tagger,M.,\& Yamaoka, K. 2002, A\&A, 386, 271
\bibitem[Rodriguez et al. (2008)]{ApJ..675..1449} Rodriguez, J., Shaw, S. E., Hannikainen, D. C., et al. 2008, ApJ, 675, 1449
\bibitem[SS(1973)]{AA..24..337} Shakura N. I., Sunyaev R. A., 1973, A\&A, 24, 337
\bibitem[Shang et al. (2019)]{Shang19} Shang J. R., Debnath D., Chatterjee D., Jana A., Chakrabarti S. K. et al., 2019, \apj, 875, 4
\bibitem[Singh et al.(2017)]{JApA..38..29} Singh, K. P., Stewart, G. C., Westergaard, N. J., et al. 2017, JApA, 38, 29
\bibitem[Stella (1999)]{PRL...82..17} Stella L., Vietri M., 1999, Physical Review Letters, 82, 17
\bibitem[ST (1985)]{AA..143..374} Sunyaev R. A., Titarchuk L. G., 1985, A\&A, 143, 374
\bibitem[Tagger (1999)]{AA..349..1003} Tagger M., Pellat R., 1999, A\&A, 349, 1003
\bibitem[Tanaka (1995)]{Cambridge Univ. Press} Tanaka, Y., \& Lewin, W. 1995, in X-Ray Binaries, Eds. W. Lewin, J. van Paradijs, \& P. van den Heuvel, Cambridge Univ. Press, Cambridge, 126
\bibitem[Vadawale et al.(2001)]{AA..370..17-21} Vadawale, S.V., Rao, A.R., Nandi, A. \& Chakrabarti, S.K., 2001, A\&A 370, 17-21
\bibitem[Vadawale et al.(2003)]{ApJ..597...1023} Vadawale, S. V., Rao, A. R., Naik, S., Yadav, J. S., Ishwara-Chandra, C. H., Pramesh Rao, A., \& Pooley, G. G. 2003, ApJ, 597, 1023
\bibitem[Vadawale et al.(2015)]{AA..578..A73} Vadawale, S. V., Chattopadhyay, T., Rao, A. R. et al., 2015, Astron. Astrophys., 578, A73.
\bibitem[Yadav et al.(2016a)]{Proc. SPIE..9905..99051D} Yadav, J. S., Agrawal, P. C., Antia, H. M., et al. 2016a, Proc. SPIE, 9905, 99051D
\bibitem[Yadav et al.(2016b)]{ApJ..833..27} Yadav, J. S., Mishra, R., Chauhan, J. V., et al. 2016b, ApJ, 833, 27


\end{thebibliography}
\end{document}